\documentclass[journal=jacsat,manuscript=article]{achemso}

\usepackage[version=3]{mhchem} % Formula subscripts using \ce{}
\usepackage[]{mathrsfs}
\usepackage[]{listings}
\usepackage[latin1]{inputenc}
\usepackage{amssymb}
\usepackage{epsfig}
\usepackage{xspace}
\usepackage{pstricks}
\usepackage{psfrag}
\usepackage{textgreek}

\newcounter{myitem}

\newcommand{\resetmycnt}{\setcounter{myitem}{1}}

\resetmycnt

\newcommand{\figdir}  {./figures/paper}

\author{\vspace*{-0.1cm}\it Yury Yu. Illarionov}
\affiliation{Southern University of Science and Technology, 1088 Xueyuan Blvd, 518055 Shenzhen, China}
\alsoaffiliation{Institute for Microelectronics (TU Wien), Gusshausstrasse 27--29, 1040 Vienna, Austria}
\email{illarionov@sustech.edu.cn}
\alsoaffiliation{Ioffe Institute, Polytechnicheskaya 26, 194021 St-Petersburg, Russia}
\author{\vspace*{-0.1cm}\it Theresia Knobloch}
\affiliation{Institute for Microelectronics (TU Wien), Gusshausstrasse 27--29, 1040 Vienna, Austria}
\author{\vspace*{-0.1cm}\it Burkay Uzlu}
\affiliation{AMO GmbH, Otto-Blumenthal-Strasse 25, 52074 Aachen, Germany}
\author{\vspace*{-0.1cm}\it Alexander G. Banshchikov}
\affiliation{Ioffe Institute, Polytechnicheskaya 26, 194021 St-Petersburg, Russia}
\author{\vspace*{-0.1cm}\it Iliya A. Ivanov}
\affiliation{Ioffe Institute, Polytechnicheskaya 26, 194021 St-Petersburg, Russia}
\author{\vspace*{-0.1cm}\it Viktor Sverdlov}
\affiliation{Institute for Microelectronics (TU Wien), Gusshausstrasse 27--29, 1040 Vienna, Austria}
\author{\vspace*{-0.1cm}\it Mikhail I. Vexler}
\affiliation{Ioffe Institute, Polytechnicheskaya 26, 194021 St-Petersburg, Russia}
\author{\vspace*{-0.1cm}\it Michael Waltl}
\affiliation{Institute for Microelectronics (TU Wien), Gusshausstrasse 27--29, 1040 Vienna, Austria}
\author{\vspace*{-0.1cm}\it Zhenxing Wang}
\affiliation{AMO GmbH, Otto-Blumenthal-Strasse 25, 52074 Aachen, Germany}
\author{\vspace*{-0.1cm}\it Bibhas Manna}
\affiliation{Institute for Microelectronics (TU Wien), Gusshausstrasse 27--29, 1040 Vienna, Austria}
\author{\vspace*{-0.1cm}\it Daniel Neumaier}
\affiliation{AMO GmbH, Otto-Blumenthal-Strasse 25, 52074 Aachen, Germany}
\alsoaffiliation{Bergische Universitat Wuppertal, Gaussstrasse 20, 42119 Wuppertal, Germany}
\author{\vspace*{-0.1cm}\it Max Christian Lemme}
\affiliation{AMO GmbH, Otto-Blumenthal-Strasse 25, 52074 Aachen, Germany}
\alsoaffiliation{RWTH Aachen University, Otto-Blumenthal-Strasse 25, 52074 Aachen, Germany}
\author{\vspace*{-0.1cm}\it Nikolai S. Sokolov}
\affiliation{Ioffe Institute, Polytechnicheskaya 26, 194021 St-Petersburg, Russia}
\author{\vspace*{-0.1cm}\it Tibor Grasser}
\affiliation{Institute for Microelectronics (TU Wien), Gusshausstrasse 27--29, 1040 Vienna, Austria}
\email{grasser@iue.tuwien.ac.at}

\title[An \textsf{achemso} demo]
  {Variability and High Temperature Reliability of Graphene Field-Effect Transistors with Thin Epitaxial CaF$_2$ Insulators}

\begin{document}

%%%%%%%%%%%%%%%%%%%%%%%%%%%%%%%%%%%%%%%%%%%%%%%%%%%%%%%%%%%%%%%%%%%%%
%% The "tocentry" environment can be used to create an entry for the
%% graphical table of contents. It is given here as some journals
%% require that it is printed as part of the abstract page. It will
%% be automatically moved as appropriate.
%%%%%%%%%%%%%%%%%%%%%%%%%%%%%%%%%%%%%%%%%%%%%%%%%%%%%%%%%%%%%%%%%%%%%

%%%%%%%%%%%%%%%%%%%%%%%%%%%%%%%%%%%%%%%%%%%%%%%%%%%%%%%%%%%%%%%%%%%%%
%% The abstract environment will automatically gobble the contents
%% if an abstract is not used by the target journal.
%%%%%%%%%%%%%%%%%%%%%%%%%%%%%%%%%%%%%%%%%%%%%%%%%%%%%%%%%%%%%%%%%%%%%

\begin{abstract}
 Graphene is a promising material for applications as a channel in graphene field-effect transistors (GFETs) which may be used as a building block for optoelectronics, high-frequency devices and sensors. However, these devices require gate insulators which ideally should form atomically flat interfaces with graphene and at the same time contain small densities of traps to maintain high device stability. Previously used amorphous oxides, such as SiO$_2$ and Al$_2$O$_3$, however, typically suffer from oxide dangling bonds at the interface, high surface roughness and numerous border oxide traps. In order to address these challenges, here we use for the first time 2$\,$nm thick epitaxial CaF$_2$ as a gate insulator in GFETs. By analyzing device-to-device variability for over 200 devices fabricated in two batches, we find that tens of them show similar gate transfer characteristics. Our statistical analysis of the hysteresis up to 175$^{\mathrm{o}}$C has revealed that while an ambient-sensitive counterclockwise hysteresis can be present in some devices, the dominant mechanism is thermally activated charge trapping by border defects in CaF$_2$ which results in the conventional clockwise hysteresis. We demonstrate that both the hysteresis and bias-temperature instabilities in our GFETs with CaF$_2$ are comparable to similar devices with SiO$_2$ and Al$_2$O$_3$. In particular, we achieve a small hysteresis below 0.01$\,$V for equivalent oxide thickness (EOT) of about 1$\,$nm at the electric fields up to 15$\,$MV/cm and sweep times in the kilosecond range. Thus, our results demonstrate that crystalline CaF$_2$ is a promising insulator for highly-stable GFETs. % despite higher applied gate fields and enhanced gate control obtained with CaF$_2$
\end{abstract}
%\textbf{Keywords:} Black phosphorus, MoS$_2$, graphene, transistor, 2D electronics, reliability improvement
%Fabrication of different electronic devices with atomically thin two-dimensional (2D) layers has achieved a considerable progress in the past few years. In particular, these emerging technologies have been 
%suggested for applications in flexible electronics~\cite{LEE13,AKINWANDE14}, optoelectronic devices~\cite{AVOURIS14,BABLICH16} and sensors~\cite{SHAVANOVA16}. Also,
%Graphical abstract
%\begin{figure}[!h]
%\vspace{0mm}
%\begin{minipage} {\textwidth} %{\sminipagewidth}
%\hspace{2cm}
 % \includegraphics[width=12cm]{\figdir/graa-eps-converted-to.pdf} %???
%\end{minipage}
%\end{figure}
%\begin{figure}[!h]
%\vspace{0mm}
%\begin{minipage} {\textwidth} %{\sminipagewidth}
%\hspace{4.5cm}
%  \includegraphics[width=9cm]{\figdir/graa-eps-converted-to.pdf} %???
%\end{minipage}
%\end{figure}
%%%%%%%%%%%%%%%%%%%%%%%%%%%%%%%%%%%%%%%%%%%%%%%%%%%%%%%%%%%%%%%%%%%%%
%% Start the main part of the manuscript here.
%%%%%%%%%%%%%%%%%%%%%%%%%%%%%%%%%%%%%%%%%%%%%%%%%%%%%%%%%%%%%%%%%%%%%

Graphene is a promising material with numerous fascinating properties~\cite{NOVOSELOV04,GEIM07} which can be attractive for applications in optoelectronics~\cite{GENG19}, sensing~\cite{BERAUD21} and radio-frequency electronics~\cite{LIAO12}. Like any other field-effect device, graphene field-effect transistors (GFETs)~\cite{LEMME07,GENG19,BERAUD21} require high-quality insulators to separate the gate from the channel. However, conventionally used amorphous oxides such as SiO$_2$, Al$_2$O$_3$ and HfO$_2$ form ill-defined interfaces with 2D materials which degrade the mobility and contain numerous border traps~\cite{FLEETWOOD92} which cause severe hysteresis~\cite{WANG10,HE16} and long-term drifts of the gate transfer characteristics~\cite{ILLARIONOV20A}. As of now, the only alternative gate insulator which has been used in GFETS is hBN which enables high mobility \cite{GANNETT11,PETRONE12,FAZIO19}. However, synthesis of high-quality hBN films on large-area substrates typically requires temperatures of more than 800$^{\mathrm{o}}$C~\cite{HUI18} which does not match the thermal budget of CMOS technologies. As a result, despite the progress already achieved in the technologies of graphene devices, the lack of suitable insulators is a central obstacle for the production of commercially competitive GFETs which will complicate their use, for instance, in currently discussed CMOS-X circuits coupling Si and 2D elements~\cite{LEMME22}.

As a promising alternative to amorphous oxides and hBN, here we use for the first time 2$\,$nm thick epitaxial calcium fluoride (fluorite, CaF$_2$) as a gate insulator in scalable GFETs with a graphene channel grown by chemical vapour deposition (CVD) and transferred onto the CaF$_2$ substrate. CaF$_2$ is an ionic crystalline insulator with good dielectric properties ($E_{\mathrm{g}}\,=\,12.1\,$eV, $\varepsilon$ in range between 6.8~\cite{HARTNETT04} and 8.43~\cite{HAYES74}) which forms quasi van der Waals interfaces with 2D materials~\cite{KOMA90} and at the same time can be epitaxially grown on Si(111) at 250$^{\mathrm{o}}$C~\cite{ILLARIONOV14D} in line with CMOS thermal budget requirements. This, in particular, makes CaF$_2$ an attractive candidate for the gate insulator of 2D FETs, even more so as CaF$_2$ allows the heteroepitaxy of 2D semiconductors on CaF$_2$(111), as already confirmed for MoSe$_2$~\cite{VISHWANATH15} and MoTe$_2$~\cite{VISHWANATH18}. In our recent works~\cite{ILLARIONOV19A,ILLARIONOV20Z} we have used CaF$_2$ to fabricate MoS$_2$ FETs with equivalent oxide thicknesses (EOT) down to 1$\,$nm and with promising performance characteristics, such as a subthreshold swing (SS) down to 90$\,$mV/dec, on/off current ratios up to 10$^7$ and high stability with respect to hysteresis and long-term drifts of the gate transfer characteristics. Recently it has been also shown that CaF$_2$ can be epitaxially grown on silicene~\cite{NAZZARI22} which opens the path towards future top gate integration of 2D materials.

Thus, as the next step in this work we extend our previous findings towards More than Moore electronics based on 2D materials which suffers from similar problems of forming high quality interfaces with insulators with few charge traps~\cite{ILLARIONOV20A}. We open a way to the further development of scalable GFETs with various fluoride materials not limited to CaF$_2$ but including also MgF$_2$ or SrF$_2$~\cite{RAVEZ97,KAVEEV05,BANSCHIKOV15} and attempt to estimate the real potential of CaF$_2$/graphene technologies by benchmarking the device-to-device variability, hysteresis and bias-temperature instabilities (BTI) of the gate transfer characteristics. We examine more than 200 GFETs with different channel dimensions and study the hysteresis and BTI dynamics in these devices for a broad range of temperatures from 25$^{\mathrm{o}}$C to 175$^{\mathrm{o}}$C. After minimizing the impact of non-insulator defects by annealing at 175$^{\mathrm{o}}$C, we demonstrate that the stable clockwise hysteresis as well as the BTI drifts in our GFETs with CaF$_2$ are comparable to those in GFETs and MoS$_2$ FETs with SiO$_2$ and Al$_2$O$_3$, despite being subjected to higher gate bias stresses. It is worth noting that the use of thin insulators allows to achieve gate fields of up to 15$\,$MV/cm which is higher than in most previously studied devices with 2D channels. This constitutes the worst case scenario in terms of gate bias stress and thus makes the small observed degradation more valuable. Therefore, we conclude that CaF$_2$ is a promising insulator for next-generation graphene technologies, including Hall sensors for high temperature applications~\cite{PETERS19} which would benefit from stable behavior of our GFETs at least up to 175$^{\mathrm{o}}$C. Furthermore, by using just 2$\,$nm thick CaF$_2$ layers we for the first time achieve CMOS-compatible gate voltage operation ranges of only several Volts for our GFETs, while also reducing the power consumption and improving the sensitivity.

Although our first proof-of-concept GFETs with CaF$_2$ do not yet offer outstanding performance in terms of mobility, we demonstrate excellent device stability and reliability owing to the crystalline CaF$_2$ grown at only 250$^{\mathrm{o}}$C. These results should boost future research on direct CVD growth of graphene on CaF$_2$ and adaptation of more mature dry transfer methods of 2D films~\cite{ASSELBERGHS20} to achieve the formation of high quality quasi-van der Waals interfaces with CaF$_2$ (111) while avoiding the ubiquitous polymer contamination from the transfer process~\cite{TILMANN2023}.

\section{CaF$_2$/Graphene Devices}

Our devices are single-layer back-gated GFETs fabricated by conventional photolithography on Si/CaF$_2$ substrates. Thin layers of CaF$_2$ (2$\,$nm, EOT$\,\sim\,$1$\,$nm) were grown on moderately doped ($N_{\mathrm{D}}\,=\,10^{15}\,$cm$^{-3}$) and highly doped ($N_{\mathrm{D}}\,=\,5\,\times\,10^{18}\,$cm$^{-3}$) n-Si(111) substrates (Batch\#1 and Batch\#2, respectively) using an established molecular beam epitaxy method at a growth temperature of 250$^{\mathrm{o}}$C~\cite{ILLARIONOV14D}, similar to our previous works on CaF$_2$/MoS$_2$ FETs~\cite{ILLARIONOV19A,ILLARIONOV20Z}. To avoid leakage currents from the large contact pads, they have been isolated with 10$\,$nm Al$_2$O$_3$ layers grown by plasma enhanced atomic layer deposition before sputtering 25$\,$nm Pd source and drain contacts. Next, a commercial CVD-grown graphene film was transferred onto the substrate using a PMMA assisted transfer method and patterned with an oxygen plasma. More details about the CaF$_2$ growth and fabrication process of our GFETs can be found in the Methods section.

The schematic layout of our GFETs is shown in Fig.1a. The obtained arrays contain hundreds of devices with channel dimensions ($L$$\,\times$$W$) from 160$\,\mu$m$\,\times$100$\,\mu$m down to 9$\,\mu$m$\,\times$3$\,\mu$m, the optical images of the device structures with different dimensions are shown in Fig.1b,c,d. In Fig.1e we show that the gate leakage current is negligible as compared to the drain current while also being far below the density of 1$\,$A/cm$^2$ at $V_{\mathrm{G}}\,=\,1\,$V, a guideline for scaled devices~\cite{ROBERTSON04}. As demonstrated in Fig.1f, typical $I_{\mathrm{D}}$-$V_{\mathrm{G}}$ characteristics of our devices with $L\,\times W\,=\,$80$\,\mu$m$\,\times$50$\,\mu$m exhibit relatively high currents up to 32$\,\mu$A/$\mu$m within few Volts operation range due to the highly downscaled thickness of the gate insulator to only 2$\,$nm.  At the same time, the $I_{\mathrm{D}}$-$V_{\mathrm{D}}$ characteristics presented in Fig.1g show good current control with some kinks typical for ambipolar GFETs~\cite{ILLARIONOV14B}. Using a similar fabrication process but without isolating the contact pads, we also fabricated similar back-gated GFETs on highly doped Si substrates with 90$\,$nm SiO$_2$ and 36$\,$nm Al$_2$O$_3$ and used them as a reference when comparing the obtained results.

\begin{figure}[!h]
\vspace{0mm}
\begin{minipage} {\textwidth} %{\sminipagewidth}
\hspace{0.2cm}
  \includegraphics[width=16cm]{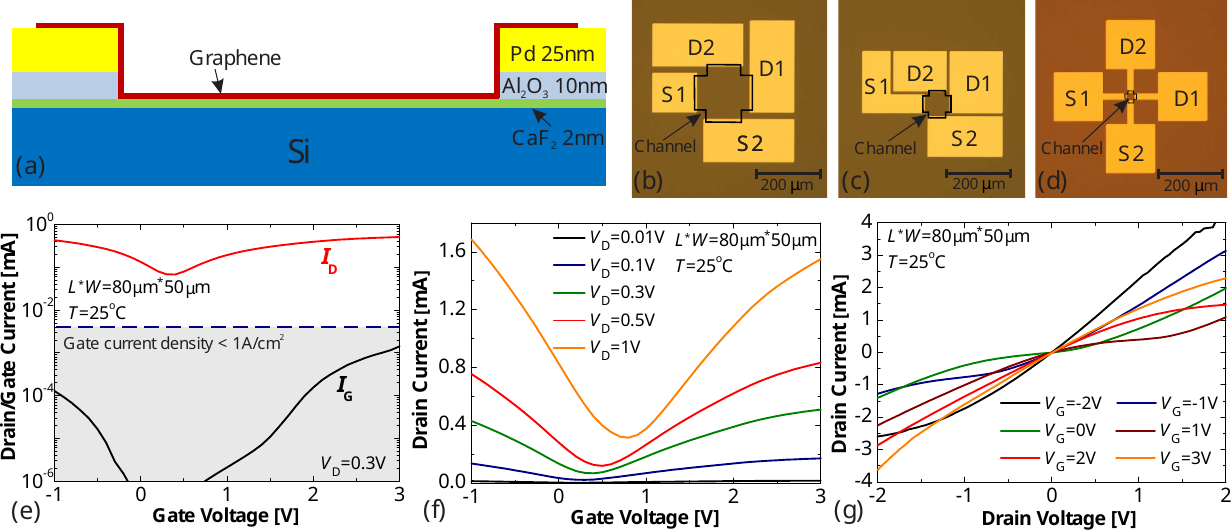} %???
\caption{\label{Fig.1} (a) Schematic structure of our back-gated GFETs with 2$\,$nm CaF$_2$ insulators. Optical images of GFETs with $L\,\times W$ of 160$\,\mu$m$\,\times$100$\,\mu$m (b), 80$\,\mu$m$\,\times$50$\,\mu$m (c) and smaller dimensions from 40$\,\mu$m$\,\times$20$\,\mu$m to 9$\,\mu$m$\,\times$3$\,\mu$m (d). (e) The gate leakage current through our thin CaF$_2$ layers is small compared to the drain current through the GFET channel with a density far below 1$\,$A/cm$^2$ at $V_{\mathrm{G}}\,=\,1\,$V. (f) Typical $I_{\mathrm{D}}$-$V_{\mathrm{G}}$ characteristics of our GFETs measured at different drain voltages. (g) The $I_{\mathrm{D}}$-$V_{\mathrm{D}}$ characteristics measured for the same GFET at different gate voltages exhibit ambipolar kinks. All provided results have been obtained for Batch\#1 GFETs.}
\end{minipage}
\end{figure}

We also note that since this is the first proof of concept study of GFETs with CaF$_2$ which employs transferred graphene films and non-protected channels, we did not focus on achieving the highest possible field-effect mobilities. Thus, the typical mobility measured using four-probe measurements does not exceed 1000$\,$cm$^2$/Vs which is low compared to most previous studies for GFETs with SiO$_2$~\cite{GANNETT11} and hBN~\cite{GANNETT11,PETRONE12,FAZIO19}. However, it is expected that a more carefully adjusted device fabrication process could result in considerably improved mobilities owing to the quasi van der Waals nature of the CaF$_2$/graphene interface. In particular, possible cracks of the already transferred graphene films due to the chemical inertness of CaF$_2$(111) surfaces~\cite{KOMA90} have to be avoided when developing specific transfer methods for this type of substrates. Furthermore, long ambient exposure of the CaF$_2$ surface prior to fabrication of GFETs was unavoidable for these prototypes but should be minimized in the future.

\section{Device-to-device variability}

We start with a statistical analysis of the $I_{\mathrm{D}}$-$V_{\mathrm{G}}$ characteristics measured for our GFETs with different sizes. As shown in Fig.2a for our Batch\#1 GFETs fabricated on moderately doped Si, already at this early stage of research over 30$\%$ of devices from the most representative group of GFETs with 80$\,\mu$m$\,\times$50$\,\mu$m dimensions exhibit very similar or even nearly identical $I_{\mathrm{D}}$-$V_{\mathrm{G}}$ characteristics, even though in overall the device-to-device variability is still sizable (see Fig.S1 in the Supporting Information (SI)). In Fig.2b we show the distribution of the Dirac current vs. Dirac voltage points ($I_{\mathrm{Dirac}}$ vs. $V_{\mathrm{Dirac}}$) for all 116 studied devices and note that the variability is stronger for GFETs with smaller channels. This is likely because our CVD-grown graphene is polycrystalline and thus larger devices may contain several complete grains within the channel, while the channel area of their smaller counterparts can be affected considerably by the grain boundaries. Since these grain boundaries significantly affect the electrostatics and carrier transport in the channel, broader distributions of $I_{\mathrm{Dirac}}$ and $V_{\mathrm{Dirac}}$ for smaller devices are to be expected. Furthermore, the same argumentation holds true for local imperfections and charges at the CaF$_2$ surface. In general, it is expected that there are microscopic inhomogeneities in the grown CaF$_2$ layers and impurities that have attached to the CaF$_2$ surface during transport and processing before the graphene layer transfer. These atomic defects will have a more pronounced impact on the charge transport for small area devices as compared to larger area ones~\cite{ASENOV03}. Another factor contributing to the variability could be different contact resistances of the pads due to some imperfections in their processing (see Fig.S2 in the SI). Thus, we expect that by further optimizing the CVD growth of the graphene channel and the device fabrication techniques, as well as the CaF$_2$ growth and the overall device fabrication flow, it may be possible to considerably reduce this variability.

\begin{figure}[!h]
\vspace{0mm}
\begin{minipage} {\textwidth} %{\sminipagewidth}
\hspace{0.2cm}
  \includegraphics[width=16cm]{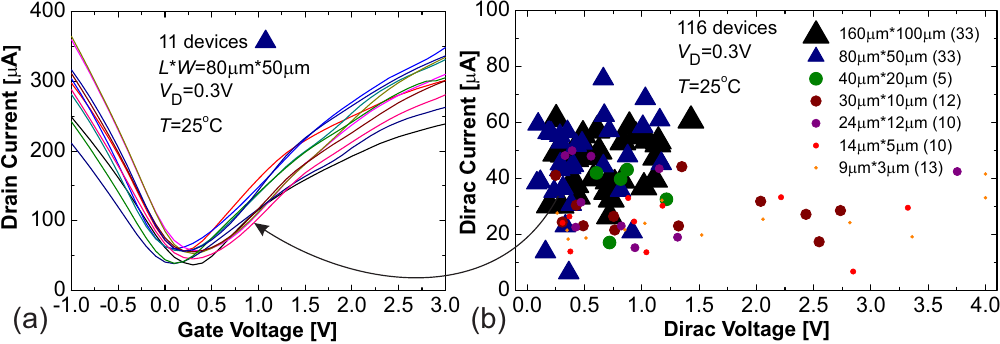} %???
\caption{\label{Fig.2} (a) $I_{\mathrm{D}}$-$V_{\mathrm{G}}$ characteristics of 11 similar GFETs with $80\mu$m$\,\times$$50\mu$m channels selected from our total statistics of 116 Batch\#1 devices with different channel dimensions. (b) Distribution of $I_{\mathrm{Dirac}}$ vs. $V_{\mathrm{Dirac}}$ for all 116 devices; the number of devices with the corresponding channel dimension is marked in brackets in the legend. The measurements have been performed before any annealing step and some smaller devices have a more positive $V_{\mathrm{Dirac}}$ (for those GFETs we used a $V_{\mathrm{G}}$ sweep range from 0 to 4$\,$V), implying the existence of a significant amount of negative charge at the interface.}
\end{minipage}
\end{figure}

In Fig.S3 in the SI we also show that GFETs from Batch\#2 which we have fabricated later on highly doped Si substrates have very similar $I_{\mathrm{D}}$-$V_{\mathrm{G}}$ characteristics and variability trends. Unlike Batch\#1 devices, before the measurements they have been subjected to initial annealing consisting of 2 days at 100$^{\mathrm{o}}$C and 5 hours at 175$^{\mathrm{o}}$C. At the same time, our reference GFETs with 36$\,$nm Al$_2$O$_3$ insulators typically exhibit larger variability in $V_{\mathrm{Dirac}}$ but lower variations in drain current (Fig.S4 in the SI). This could hint at a higher quality of the graphene films after their transfer onto a non-inert Al$_2$O$_3$ surface but at the same time larger and more variable number of fixed charges at the graphene/Al$_2$O$_3$ interface as compared to graphene/CaF$_2$.

\begin{figure}[!h]
\hspace{0cm}
\vspace{-3mm}
\begin{minipage} {\textwidth} %{\sminipagewidth}
\hspace{2cm}
  \includegraphics[width=14cm]{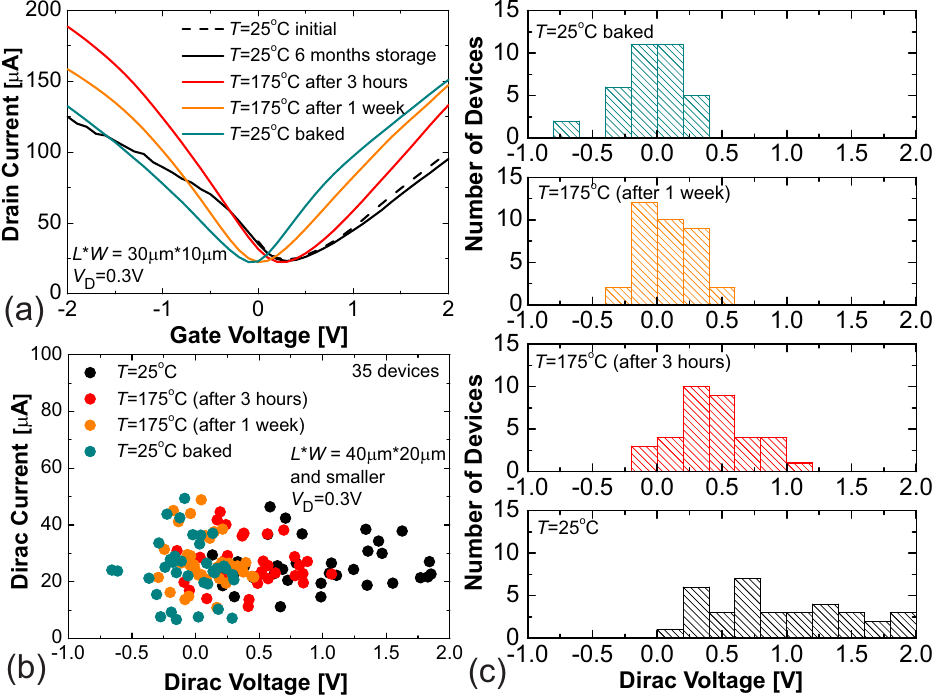} %???
\caption{\label{Fig.3} (a) Transformation of the $I_{\mathrm{D}}$-$V_{\mathrm{G}}$ characteristics following 6 months of storage under a moderate vacuum, subsequent annealing at 175$^{\mathrm{o}}$C and cooling back to 25$^{\mathrm{o}}$C. (b) Distribution of $I_{\mathrm{Dirac}}$ vs. $V_{\mathrm{Dirac}}$ for 35 devices at 25$^{\mathrm{o}}$C, in the beginning and in the end of 175$^{\mathrm{o}}$C annealing in vacuum (10$^{-6}$$\,$torr), and at 25$^{\mathrm{o}}$C after annealing. (c) Statistical distributions of $V_{\mathrm{Dirac}}$ values for these devices at the same temperature/annealing conditions and at $V_{\mathrm{D}}\,=\,0.3\,$V.} 
\vspace*{1cm}
\end{minipage}
\end{figure}

In addition, we have repeated the variability measurements on our Batch\#1 GFETs following 6 months of storage under a moderate vacuum of about 600$\,$torr. We found that at least 35 out of 116 GFETs remained functional and thus analyzed the device-to-device variability under the impact of elevated temperatures, while ignoring about 10$\,\%$ of outliers with too positive $V_{\mathrm{Dirac}}$ above 2$\,$V (Fig.2b) in our statistics. As shown in Fig.3a for one representative GFET, at room temperature the measured $I_{\mathrm{D}}$-$V_{\mathrm{G}}$ characteristic does not change significantly following long storage. However, annealing at 175$^{\mathrm{o}}$C results in a negative shift of $V_{\mathrm{Dirac}}$ which becomes more pronounced after a week at high temperature and does not recover after cooling back to 25$^{\mathrm{o}}$C. This is likely caused by evaporation of some impurities from graphene or adsorbates which could serve as fixed charges and affect $V_{\mathrm{Dirac}}$. The $I_{\mathrm{Dirac}}$ vs. $V_{\mathrm{Dirac}}$ distributions obtained for 35 devices with channel dimensions ranging from 40$\,\mu$m$\,\times$20$\,\mu$m down to 9$\,\mu$m$\,\times$3$\,\mu$m (Fig.3b) suggest that this negative drift of $V_{\mathrm{Dirac}}$ is a common feature for all GFETs on CaF$_2$. Furthermore, the variability in $V_{\mathrm{Dirac}}$ becomes smaller after annealing, as also confirmed by the statistical distributions shown in Fig.3c. Thus, we suggest that a thermal treatment of the devices at 175$^{\mathrm{o}}$C should allow to exclude side effects related to the ambient impact on our GFETs, thus revealing the hysteresis dynamics which could be attributed solely to border insulator defects in CaF$_2$.

\section{Hysteresis dynamics and reliability}

In Fig.4 we analyze the hysteresis dynamics in GFETs with 80$\,\mu$m$\,\times$50$\,\mu$m channels. Among five selected devices, four are from Batch\#1 which have not experienced any annealing and one is from Batch\#2 which has been subjected only to initial annealing, i.e. 2 days at 100$^{\mathrm{o}}$C followed by 5 hours at 175$^{\mathrm{o}}$C. We can see that all devices have similar $I_{\mathrm{D}}$-$V_{\mathrm{G}}$ characteristics which confirms the good reproducibility of our GFET technology. The hysteresis width ($\Delta$$V_{\mathrm{H}}$) vs. reciprocal sweep time (1/$t_{\mathrm{sw}}$) dependencies~\cite{ILLARIONOV16A} measured for these GFETs are shown in Fig.4b. While all devices from Batch\#1 exhibit switching of the hysteresis from counterclockwise at faster sweeps to clockwise at slower sweeps, a typical GFET from Batch\#2 has only a small clockwise hysteresis. In Fig.S5 in the SI we show the $I_{\mathrm{D}}$-$V_{\mathrm{G}}$ characteristics of these devices measured using different sweep rates and observe that the clockwise hysteresis which appears for slow sweeps for Batch\#1 GFETs is accompanied with a permanent negative drift of $V_{\mathrm{Dirac}}$. We suggest that while the conventional clockwise hysteresis is caused by fast insulator defects located close to the CaF$_2$/graphene interface, the permanent drift accumulated during multiple sweeps is similar to bias-temperature instabilities (BTI) known from Si technologies~\cite{GRASSER11A}, thus being a consequence of the charging of slower insulator defects in CaF$_2$ which have time constants in the range of kiloseconds. Remarkably, for the GFETs from Batch\#2 both the hysteresis and negative drift of $V_{\mathrm{Dirac}}$ are considerably less pronounced, which could mean that the work function of graphene used in our second fabrication round is more favorable to energetically suppress the charge trapping by defects in CaF$_2$~\cite{KNOBLOCH22}.

\begin{figure}[!h]
\hspace{0cm}
\vspace{-3mm}
\begin{minipage} {\textwidth} %{\sminipagewidth}
\hspace{1cm}
  \includegraphics[width=14cm]{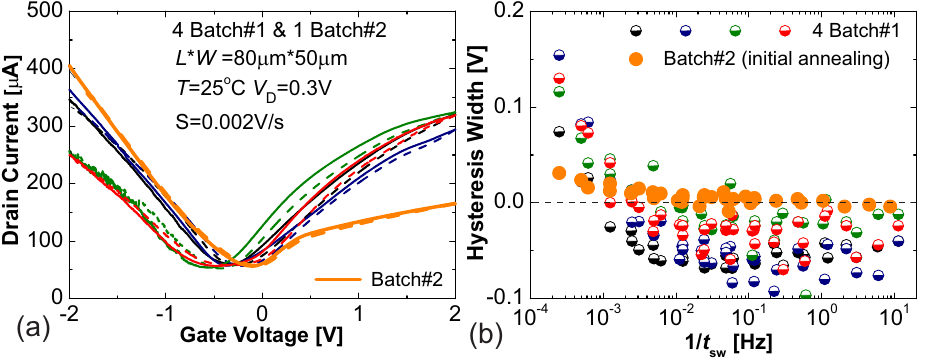} %???
\caption{\label{Fig.4} (a) Double sweep $I_{\mathrm{D}}$-$V_{\mathrm{G}}$ characteristics of five GFETs with $80\mu$m$\,\times$$50\mu$m channels measured using ultra-slow sweeps with $S\,=\,0.002\,$V/s. Among these devices, there is one GFET from Batch\#2. (b) The $\Delta$$V_{\mathrm{H}}$ vs. 1/$t_{\mathrm{sw}}$ dependencies for the same GFETs. Compared to GFETs from Batch\#1, the device from Batch\#2 has only a small clockwise hysteresis at slow sweeps with no counterclockwise hysteresis at fast sweeps. }
\vspace*{1cm}
\end{minipage}
\end{figure}

\begin{figure}[!h]
\hspace{0cm}
\vspace{-3mm}
\begin{minipage} {\textwidth} %{\sminipagewidth}
\hspace{1cm}
  \includegraphics[width=14cm]{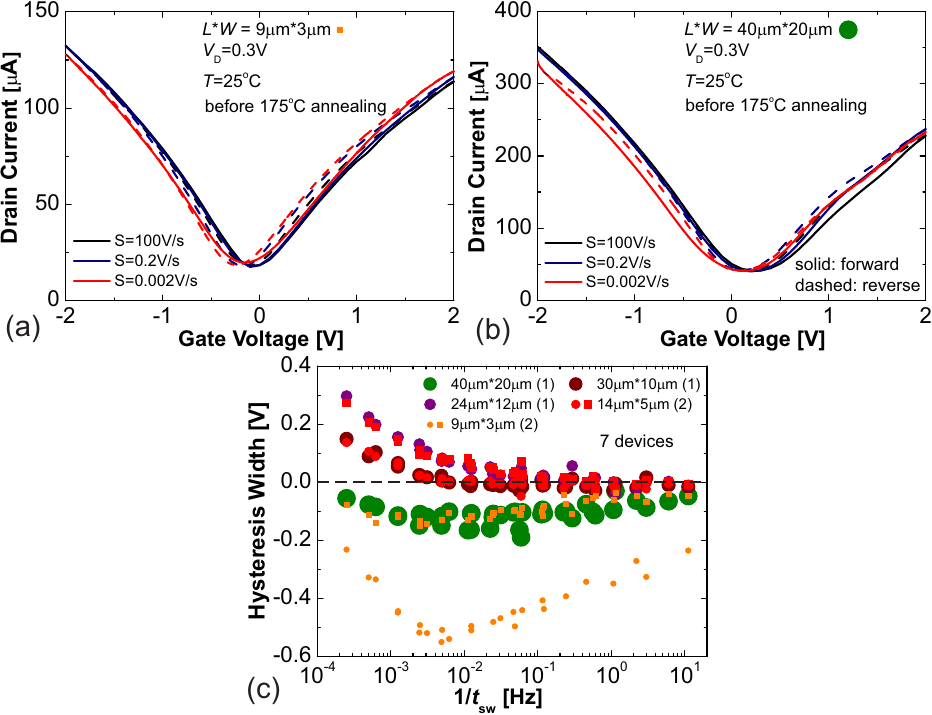} %???
\caption{\label{Fig.5} Double sweep $I_{\mathrm{D}}$-$V_{\mathrm{G}}$ characteristics of GFETs with $9\mu$m$\,\times$$3\mu$m (a) and $40\mu$m$\,\times$$20\mu$m (b) channels measured with different sweep rates. The hysteresis dynamics observed for these two devices are very similar. (c) The $\Delta$$V_{\mathrm{H}}$ vs. 1/$t_{\mathrm{sw}}$ dependencies for seven GFETs with different channel dimensions. While there is some variability in the hysteresis dynamics, some devices with different sizes have identical hysteresis and thus this effect appears to be independent of the channel dimensions. } 
\vspace*{1cm}
\end{minipage}
\end{figure}

\begin{figure}[!h]
\hspace{0cm}
\vspace{-3mm}
\begin{minipage} {\textwidth} %{\sminipagewidth}
\hspace{2cm}
  \includegraphics[width=13cm]{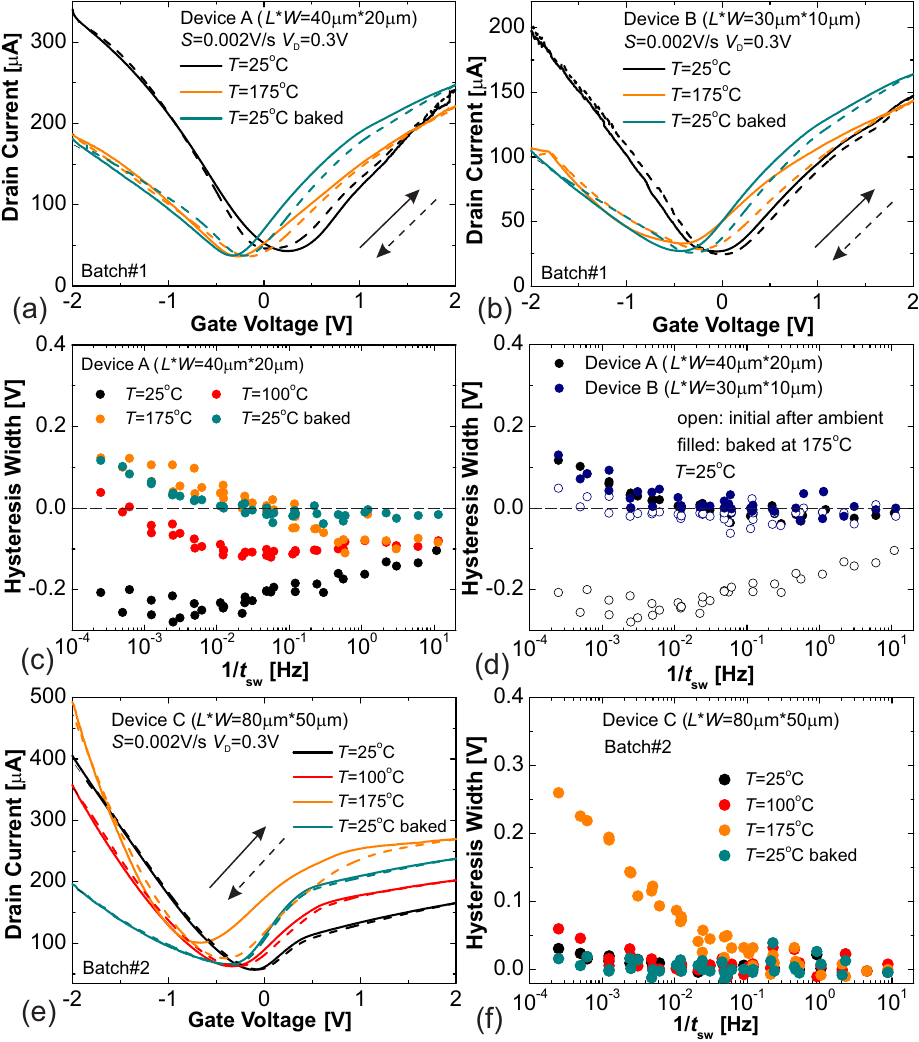} %???
\caption{\label{Fig.6} Double sweep $I_{\mathrm{D}}$-$V_{\mathrm{G}}$ characteristics measured for our Batch\#1 GFETs at $T\,=\,25^{\mathrm{o}}$C, $175^{\mathrm{o}}$C and $25^{\mathrm{o}}$C after annealing using $S\,=\,0.002\,$V/s. Just after 10 minutes of ambient exposure, Device A (a) exhibited counterclockwise and Device B (b) clockwise hysteresis. (c) At $100^{\mathrm{o}}$C and at $175^{\mathrm{o}}$C the counterclockwise hysteresis in Device A is strongly suppressed, and after annealing at $175^{\mathrm{o}}$C both devices exhibit similar clockwise hysteresis (d). (e,f) The corresponding results for a Batch\#2 GFET which show no counterclockwise contribution and conventional thermal activation of charge trapping. These GFETs experienced an initial annealing of 2 days at $100^{\mathrm{o}}$C and 5 hours at $175^{\mathrm{o}}$C prior to the first measurement round.}
\vspace*{1cm}
\end{minipage}
\end{figure}

To understand the origins of the observed hysteresis dynamics and in particular the counterclockwise hysteresis, we have performed a similar analysis on GFETs with smaller channel areas before thermal annealing. In Fig.5a,b we show the $I_{\mathrm{D}}$-$V_{\mathrm{G}}$ characteristics of Batch\#1 GFETs with $9\mu$m$\,\times$$3\mu$m and $40\mu$m$\,\times$$20\mu$m channel dimensions which exhibit similar hysteresis at different sweep rates. As confirmed in Fig.5c for a larger statistics of seven GFETs, despite the overall variability in $\Delta$$V_{\mathrm{H}}$(1/$t_{\mathrm{sw}}$) curves, some devices with different sizes show nearly identical hysteresis dynamics. Thus, this variability is not directly related to the channel dimensions but rather to the local density and type of defects near the channel. At the same time, it is remarkable that some of these smaller GFETs exhibit a counterclockwise hysteresis even at slow sweeps, while the others have purely clockwise hysteresis which becomes larger for slow sweeps and thus indicates a standard charge trapping mechanism at border traps in the CaF$_2$ close to the channel. To benchmark the origin of the counterclockwise hysteresis, we have also performed hysteresis measurements after 10 minutes of ambient exposure. As shown in Fig.S6 in the SI, in the device with initially dominant counterclockwise hysteresis this hysteresis becomes more pronounced following the ambient exposure, while the clockwise hysteresis in the second device slightly decreases without any switching to the counterclockwise direction. Therefore, we suggest that this counterclockwise hysteresis is mostly due to the interaction of our bare channel GFETs with the ambient environment while being unlikely related to defects in CaF$_2$. The possible reasons for this behavior could include, for instance, interaction of defects in graphene with adsorbates or the diffusion of oxygen through imperfections in CaF$_2$ to the Si/CaF$_2$ interface. A counterclockwise hysteresis could for example be caused by charge trapping at defects close to the gate side, formed by the Si/CaF$_2$ interface~\cite{WALDHOER22}. % We suggest that the counterclockwise hysteresis in our GFETs is due to defects at the Si/CaF$_2$ interface which can trap carriers from the global silicon back gate. Considering our previous observations of oxidation at the Si/CaF$_2$ interface and the formation of a thin SiO$_2$ layer~\cite{ILLARIONOV19A}, it is likely that while transporting the samples in an ambient exposure in between CaF$_2$ growth and GFET fabrication the density of these defects increases due to the penetration of e.g. oxygen through possible imperfections in CaF$_2$. Thus, the devices which contain some imperfections in the CaF$_2$ layer within the channel area are likely those which have an atypical counterclockwise hysteresis, while GFETs with more homogeneous CaF$_2$ should have the standard clockwise hysteresis caused by border traps at the CaF$_2$/graphene interfaces. %As these adsorbates can be trapped by defects in graphene, such as vacancies, the variability in hysteresis dynamics is likely due to more defected graphene near the grain boundaries. Thus, GFETs with more pronounced counterclockwise hysteresis should be those with more defected channels. However, for considerably larger GFETs with the channel formed by more complete grains the variability in hysteresis is not that high (Fig.4b) and the counterclockwise hysteresis is less visible, in particular at slow sweeps. %As these adsorbates can be trapped by defects in graphene, such as vacancies, the variability in hysteresis dynamics is likely due to more defected graphene near the grain boundaries.

In this context, we next analyze the impact of high temperature annealing on the hysteresis in our GFETs directly after ambient exposure. In Fig.6a,b we show the $I_{\mathrm{D}}$-$V_{\mathrm{G}}$ characteristics of two GFETs with counterclockwise (Device A) and clockwise (Device B) hysteresis at different temperatures up to 175$^{\mathrm{o}}$C and back at 25$^{\mathrm{o}}$C after six days annealing required to complete our measurements at 175$^{\mathrm{o}}$C. Indeed, the counterclockwise hysteresis in Device A can be considerably suppressed (Fig.6c), which makes the initially different $\Delta$$V_{\mathrm{H}}$(1/$t_{\mathrm{sw}}$) traces of two GFETs nearly identical after annealing (Fig.6d). Furthermore, in Batch\#2 GFETs (Fig.6e,f) we do not see any counterclockwise hysteresis and observe only a conventional thermal activation of clockwise hysteresis which is consistent with our previous findings about charge trapping by border insulator defects situated near the interface with 2D channels~\cite{ILLARIONOV16A,ILLARIONOV16B}. Remarkably, after about one week at 175$^{\mathrm{o}}$C the hysteresis in these GFETs remains small and the dynamics observed at 25$^{\mathrm{o}}$C do not change. This suggests that the density of border defects in CaF$_2$ is relatively low and that there is no thermally induced creation of new defects. Also, we note that the clockwise hysteresis in Batch\#2 GFETs is smaller as compared to their Batch\#1 counterparts, which could be explained by a more favorable work function~\cite{KNOBLOCH22} of the graphene films used during the second fabrication round. %Thus, we can conclude that charge trapping by border defects in CaF$_2$ situated close to the interface with graphene causes the conventional clockwise hysteresis observed after annealing

Next we perform BTI measurements for the Batch\#2 GFET studied in Fig.6e,f using increased gate bias stress $V_{\mathrm{G,stress}}$, constant stress time $t_{\mathrm{s}}\,=\,10\,$ks and recovery voltage $V_{\mathrm{G,rec}}\,=-0.4\,$V. The results obtained after 1 week of annealing at 175$^{\mathrm{o}}$C are shown in Fig.7. It is clear that despite extremely high insulator fields up to 15$\,$MV/cm, both NBTI (Fig.7a, $V_{\mathrm{G,stress}}<0$V)  and PBTI (Fig.7b, $V_{\mathrm{G,stress}}>0$V) drifts are comparatively small. At the same time, no anomalous trends are present. This is in line with a small clockwise hysteresis measured for the same device which is actually a superposition of NBTI and PBTI accumulated during the sweeps~\cite{ILLARIONOV17A} and again confirms low density of border traps in CaF$_2$. At the same time, the recovery of the observed degradation is rather weak in both cases, which suggests contributions from deep trap levels in CaF$_2$. This is in line with our first principle calculations for possible defects in CaF$_2$ which could be Si interstitials (Si$_{\mathrm{i}}$) or Si substituting Ca (Si$_{\mathrm{Ca}}$) energetically aligned deep in the bandgap of CaF$_2$~\cite{WALDHOER22}.

\begin{figure}[!h]
\hspace{0cm}
\vspace{-3mm}
\begin{minipage} {\textwidth} %{\sminipagewidth}
\hspace{1.5cm}
  \includegraphics[width=13cm]{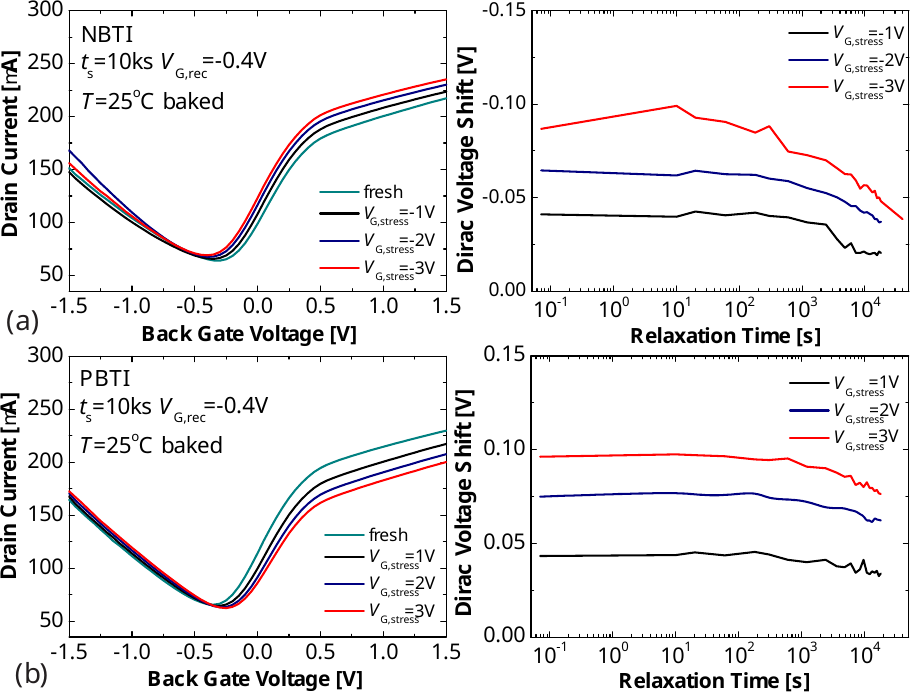} %???
\caption{\label{Fig.7} (a) Evolution of the $I_{\mathrm{D}}$-$V_{\mathrm{G}}$ under NBTI stress measured for our Batch\#2 GFET with $L\,\times W\,=\,$80$\,\mu$m$\,\times$50$\,\mu$m (left) and the corresponding recovery traces for increasing stress biases, $V_{\mathrm{G,stress}}$  (right). (b) The corresponding results for PBTI measured on the same device.}
\vspace*{1cm}
\end{minipage}
\end{figure}

\begin{figure}[!h]
% \hspace{0cm}
% \vspace{-3mm}
% \begin{minipage} {\textwidth} %{\sminipagewidth}
% \hspace{0.1cm}
  %\includegraphics[width=16cm]{\figdir/Fig.8_new-eps-converted-to.pdf}%
  \includegraphics[width=16cm]{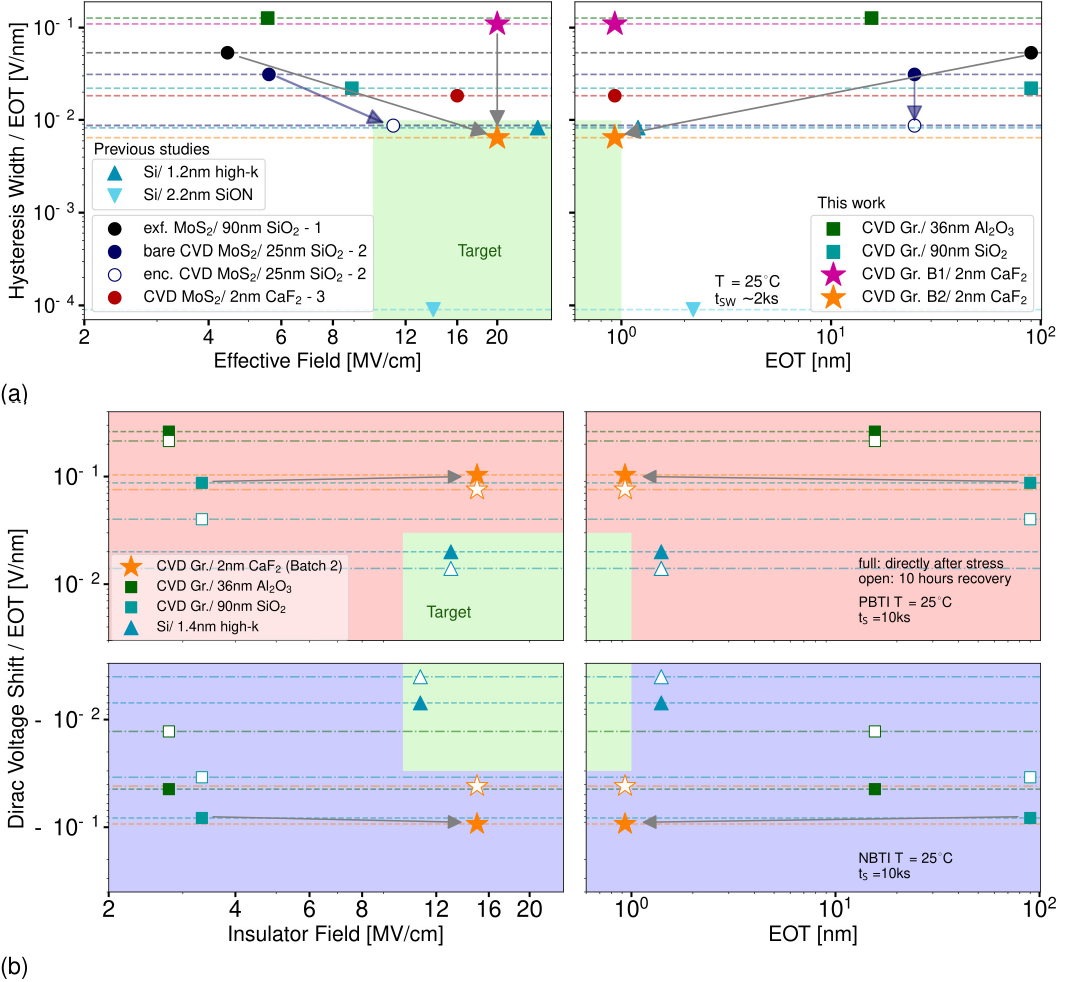}%
\caption{\label{Fig.8} (a) Comparison of the normalized hysteresis widths in different back-gated 2D FETs (\mbox{1-\cite{ILLARIONOV16A}~2-\cite{ILLARIONOV17C}~3-\cite{ILLARIONOV19A}}) and Si FETs for the sweep time of about 2$\,$ks versus the effective gate field $(V_{\mathrm{G,max}}-V_{\mathrm{G,min}})$/$d_{\mathrm{ins}}$ (left) and versus EOT (right). For comparing the hysteresis widths on technologies with different gate stacks, the hysteresis widths were normalized by EOT. The measured clockwise hysteresis in our GFETs with CaF$_2$ is comparable to the normalized hysteresis widths reported in 2D devices, with the hysteresis on our Batch\#2 CaF$_2$ GFETs being the smallest and meeting the target values. %given by the hysteresis on Si/high-k devices
(b) Comparison of normalized PBTI (top) and NBTI (bottom) drifts measured with a small time delay of about 0.5$\,$s after stress and after 10 hours of recovery versus the insulator field $V_{\mathrm{G,stress}}$/$d_{\mathrm{ins}}$ (left) and versus EOT (right). Measured Dirac shifts were normalized by EOT, revealing a comparable BTI on our GFETs with CaF$_2$ as in devices with SiO$_2$ and Al$_2$O$_3$ gate insulators, despite being stressed at much higher gate fields. } %Dashed lines (stressed) and dash-dotted line (recovered) indicate the trends of how the hysteresis width is expected to scale when scaling down the insulator thickness $d_{\mathrm{ins}}$.
%Dashed lines show how the BTI is expected to scale for thinner gate stacks $d_{\mathrm{ins}}$.
%, resulting in a quantity proportional to the number of activated charge traps, $N_\mathrm{T}$
% \vspace*{1cm}
% \end{minipage}
\end{figure}

In order to accurately compare the measured hysteresis and BTI shifts of our CaF$_2$ GFETs to the results on FETs with different gate stacks, the measured hysteresis widths $\Delta$$V_{\mathrm{H}}$ and Dirac point shifts $\Delta$$V_{\mathrm{Dirac}}$ are normalized by EOT. As both the hysteresis in the gate transfer charactersitics and BTI are caused by charge trapping, the following relation holds
\begin{equation}
 \Delta V_{\mathrm{H}} = -q \frac{N_\mathrm{T}}{C_\mathrm{ins}},
\end{equation}
with the elementary charge $q$, the density of active charge traps, $N_\mathrm{T}$, and the insulator capacitance $C_\mathrm{ins}=\varepsilon_0 \varepsilon_\mathrm{r}/d_\mathrm{ins}$. Hence, by introducing the equivalent oxide thickness (EOT=$\varepsilon_\mathrm{SiO_2}/\varepsilon_\mathrm{r} \times  d_\mathrm{ins}$) we obtain
\begin{equation}
 \frac{\Delta V_{\mathrm{H}}}{\mathrm{EOT}} = - \frac{q}{3.9 \varepsilon_0} N_\mathrm{T},
\end{equation}
with the SiO$_2$ dielectric constant of $3.9$. As a consequence, the normalized hysteresis width is directly proportional to the density of active charge traps, $N_\mathrm{T}$, with physical constants as proportionality factors.
Using this normalization, we compare in Fig.\ref{Fig.8}a the post-annealing hysteresis observed for our CaF$_2$ GFETs, for back-gated MoS$_2$ FETs with SiO$_2$~\cite{ILLARIONOV16A, ILLARIONOV17C} and CaF$_2$~\cite{ILLARIONOV19A}, for commercial silicon FETs and also for our reference back-gated GFETs with SiO$_2$ and Al$_2$O$_3$ (see more details in Fig.S7 in the SI).
$\Delta$$V_{\mathrm{H}}/\mathrm{EOT}$ measured at $t_{\mathrm{sw}}\,=\,2\,$ks is plotted versus the effective gate insulator field $(V_{\mathrm{G,max}}-V_{\mathrm{G,min}})$/$d_{\mathrm{ins}}$, where $V_{\mathrm{G,max}}$ and  $V_{\mathrm{G,min}}$ are the boundaries of the gate sweep range, and also versus EOT.
Indeed, owing to the crystalline CaF$_2$ and the highly scaled insulator thickness, the observed hysteresis is small and comparable to the normalized hysteresis observed in other devices, even at high effective gate insulator fields. This comparison shows that the density of activated charge traps $N_\mathrm{T}$ in our Batch\#2 GFETs is minimal, leading to the best observed performance.
Moreover, for the Batch\#2 GFETs the targets of $\Delta$$V_{\mathrm{H}}\,<\,0.01\,$V for EOT$\,<\,$1$\,$nm and a gate field of $\mathcal{E}_\mathrm{ins}\,>\,10\,$MV/cm are reached, showing a smaller hysteresis than measurements on a commercial Si/high-k technology.
For reference we show a similar comparison in Fig.S8 in the SI without any normalization, directly comparing the measured $\Delta$$V_{\mathrm{H}}$, demonstrating how a scaled EOT is very important to achieve small absolute numbers for the hysteresis width, $\Delta$$V_{\mathrm{H}}$ and further illustrating the excellent performance of our GFETs.

Additionally, in Fig.\ref{Fig.8}b we compare the NBTI and PBTI shifts of the Dirac voltage normalized by EOT as a function of the stress insulator field and EOT for our GFETs with CaF$_2$, Al$_2$O$_3$ and SiO$_2$ to the BTI on a commercial silicon high-k scaled logic node~\cite{Rzepa2018}. For the three GFET batches, the absolute stress voltages were 3, 10 and 30$\,$V, respectively. In agreement with the hysteresis results for GFETs with CaF$_2$, the normalized BTI shifts are comparable for all three gate insulators, even though CaF$_2$ has experienced considerably higher electric fields. In the CaF$_2$ GFETs the electric gate fields and EOT targets are achieved, even though the BTI target of $0.03\,$V according to the International Roadmap of Semiconductor Devices and Systems~\cite{IRDS2022} is out of reach for any of the compared GFET technologies. Again, in Fig.S8 in the SI a comparison of the directly measured $\Delta$$V_{\mathrm{Dirac}}$ values is shown, confirming the small BTI seen in our CaF$_2$ gated GFETs.
These results demonstrate that CaF$_2$ is a promising insulator which allows to fabricate ultra-scaled 2D devices of high stability with respect to charge trapping.

\section{Conclusions}

We fabricated over two hundred GFETs with CVD-grown graphene channels and 2$\,$nm thick epitaxial CaF$_2$ insulators and performed an in-depth study of the device-to-device variability and hysteresis dynamics. Our results show that although grain boundaries of the channel or imperfections in CaF$_2$ layers may introduce some variability in the gate transfer characteristics, some nearly identical GFETs can be found already at this early stage of research. We have also performed a comprehensive statistical analysis of hysteresis and BTI on many GFETs at temperatures of up to 175$^{\mathrm{o}}$C. Our findings suggest that the initially observed, ambient-sensitive, counterclockwise hysteresis can be fully suppressed by 175$^{\mathrm{o}}$C annealing in our first batch of GFETs and is not present in the devices from the second batch. The remaining clockwise hysteresis can be attributed to border traps in CaF$_2$ and is on devices from the second batch smaller than in reference GFETs with SiO$_2$ and Al$_2$O$_3$. The main milestone of our study is that we have achieved a hysteresis below 0.01$\,$V for an equivalent oxide thickness (EOT) of about 1$\,$nm at electric fields up to 15$\,$MV/cm and long sweep times in the kilosecond range, reaching the target values set by commercial silicon technologies. These results confirm that the use of crystalline CaF$_2$ as a gate insulator is a promising way to enable stable GFETs for sensors and optoelectronics, including Hall sensors for high temperature operations. Our findings can be generalized to include applications of various 2D materials and their devices for heterogeneous electronics coupling 2D and silicon CMOS elements~\cite{KUMAR22}.

\section{Methods}

\textit{Device fabrication} 

Fabrication of our GFETs consists of MBE growth of 2$\,$nm thick CaF$_2$ films and photolithography to produce the device arrays with channel dimensions ($L$$\,\times$$W$) from 160$\,\mu$m$\,\times$100$\,\mu$m down to 9$\,\mu$m$\,\times$3$\,\mu$m on the obtained Si/CaF$_2$ surfaces.  

CaF$_2$ layers were grown on moderately doped ($N_{\mathrm{D}}\,=\,10^{15}\,$cm$^{-3}$) and \textbf{highly doped ($N_{\mathrm{D}}\,=\,5\,\times\,10^{18}\,$cm$^{-3}$)} n-Si(111) substrates with a miscut angle of 5 to 10 minutes. Following careful chemical treatment of Si(111) surface, a protective oxide layer was formed using the method of Shiraki~\cite{ISHIZAKA86} and subsequently removed by annealing for 2 minutes at 1200$^{\mathrm{o}}$C under ultra-high vacuum conditions ($\sim$10$^{-8}$$-$10$^{-7}$$\,$Pa). After this, the CaF$_2$ film with 2$\,$nm thickness was grown on the obtained atomically clean 7$\times$7 Si(111) surface using an MBE process with the optimal growth temperature of 250$^{\mathrm{o}}$C and deposition rate of about 1.3$\,$nm/min. A crystalline quality of the obtained CaF$_2$ layers was examined in situ using reflection high-energy electron diffraction (RHEED)~\cite{SOKOLOV92} with an electron energy of 15$\,$keV. The corresponding RHEED patterns which confirm high crystallinity with single-crystal structure of our thin CaF$_2$ films can be found in the Supporting Information of our previous work~\cite{ILLARIONOV19A}.

Our GFETs were fabricated on the obtained epitaxial Si/CaF$_2$ substrates using conventional photolithography. After defining the source and drain contact regions by photolithography, 10$\,$nm Al$_2$O$_3$ was deposited by plasma enhanced atomic layer deposition to isolate the contact pads and source and drain metals were deposited by sputtering 25$\,$nm Pd, followed by a lift-off process. After fabricating the contacts, commercially available chemical vapor deposited graphene was transferred on the substrate by a PMMA assisted method. Finally, the graphene was patterned by oxygen plasma to form the transistor channel which is bottom-contacted by Pd metal. 

Using a similar approach, we have also fabricated back-gated GFETs with 90$\,$nm SiO$_2$ and 36$\,$nm Al$_2$O$_3$ insulators to be used for reference in hysteresis comparison. However, considering larger insulator thickness, no isolation of contact pads was needed in that case. 

\textit{Measurement technique} 

Electrical characterization of our GFETs with CaF$_2$ consisted in the measurements of the $I_{\mathrm{D}}-V_{\mathrm{G}}$ characteristics and hysteresis dynamics. These measurements were performed using a Keithley 2636 parameter analyzer in the chamber of a Lakeshore probestation in a vacuum ($\sim$5$\times$10$^{-6}$$\,$torr), in complete darkness and at temperatures ranging from 25$^{\mathrm{o}}$C to 175$^{\mathrm{o}}$C, with the days-long measurements at 175$^{\mathrm{o}}$C being also considered as an annealing step. The hysteresis of the $I_{\mathrm{D}}-V_{\mathrm{G}}$ characteristics was studied using our established measurement technique~\cite{ILLARIONOV16A} based on double sweeps with varied sweep times. The hysteresis width was obtained as a difference of $V_{\mathrm{Dirac}}$ between forward and reverse sweep $I_{\mathrm{D}}-V_{\mathrm{G}}$ characteristics. We express our results by plotting the hysteresis widths $\Delta$$V_{\mathrm{H}}$ versus the reciprocal sweep time $1/t_{\mathrm{sw}}$. For comparing hysteresis widths and Dirac point shifts for different devices and measurement conditions, we extract $\Delta$$V_{\mathrm{H}}$ for a sweep time in the kilosecond range, normalize it by EOT and plot it versus the effective gate field $(V_{\mathrm{G,max}}-V_{\mathrm{G,min}})/d_{\mathrm{ins}}$, where $(V_{\mathrm{G,max}}-V_{\mathrm{G,min}})$ is the width of the sweep range and $d_{\mathrm{ins}}$ is the insulator thickness, and also versus EOT.

%%%%%%%%%%%%%%%%%%%%%%%%%%%%%%%%%%%%%%%%%%%%%%%%%%%%%%%%%%%%%%%%%%%%%
%% The "Acknowledgement" section can be given in all manuscript
%% classes.  This should be given within the "acknowledgement"
%% environment, which will make the correct section or running title.
%%%%%%%%%%%%%%%%%%%%%%%%%%%%%%%%%%%%%%%%%%%%%%%%%%%%%%%%%%%%%%%%%%%%%
\begin{acknowledgement}
Useful discussions of growth experiment results with Dr. S.M.Suturin are greatly appreciated. Y.Y.I., T.K. and T.G. acknowledge financial support through the FWF grants I2606-N30, I4123-N30 (joint project with DFG LE 2440/7-1). Y.Y.I, T.K., T.G., A.G.B., I.A.I., N.S.S. and M.I.V. are also grateful to the support within the joint FWF (grant I5296-N) and RFBR (grant 21-52-14007) project. Furthermore, T.K. acknowledges financial support through the FFG under project no. 1755510. M.C.L. acknowledges financial support through the German Federal Ministry of Education and Research grant GIMMIK (03XP0210) and DFG grants ULTIMOS2 (LE 2440/7-1). Additionally, B.U., Z.W., D.N. and M.C.L. acknowledge funding from the European Union's Horizon 2020 research and innovation programme under grant agreements 2D-EPL (952792), GrapheneCore3 (881603) and WiPLASH (863337). D.N. acknowledges financial support through the DFG grant GLECSII (NE1633/3-2). M.W. acknowledges financial support by the Austrian Federal Ministry for Digital and Economic Affairs; the National Foundation for Research, Technology and Development; and the Christian Doppler Research Association. Y.Y.I. and M.I.V. acknowledge financial support by the Ministry of Science and Higher Education of the Russian Federation under project no. 075-15-2020-790 and Y.Y.I. also the start up of Southern University of Science and Technology (SUSTech).
\end{acknowledgement}

%\section{Data availability}
%The data that support the graphs within this manuscript and further details of this study are available from the corresponding author upon reasonable request. 

%\section{Author contributions} 

%Y.Y.I. introduced the idea of MoS$_2$ FETs with ultra-thin CaF$_2$ insulator, performed their characterization and prepared the manuscript. A.G.B. performed MBE growth of CaF$_2$ and provided the substrates. D.K.P. and S.W. fabricated MoS$_2$ FETs. M.S.-P. did TEM measurements. T.K. and M.T. contributed to preparation of figures. M.S-P. and A.S.-T. performed TEM measurements and sample preparation, respectively. M.I.V. performed quantitative analysis of gate leakage currents using tunnel models. M.W. programmed electrical measurements. N.S.S., T.M. and T.G. supervised this work. All authors regularly discussed the results and commented on the manuscript. 

\section{Competing interests} 
The authors declare no competing interests.

\section{Data availability} 
All data of this manuscript can be available from the authors upon request.

%%%%%%%%%%%%%%%%%%%%%%%%%%%%%%%%%%%%%%%%%%%%%%%%%%%%%%%%%%%%%%%%%%%%%
%% The appropriate \bibliography command should be placed here.
%% Notice that the class file automatically sets \bibliographystyle
%% and also names the section correctly.
%%%%%%%%%%%%%%%%%%%%%%%%%%%%%%%%%%%%%%%%%%%%%%%%%%%%%%%%%%%%%%%%%%%%%
\bibliography{./bib/diss}

\end{document}